\documentclass[11pt]{article}
\usepackage{amsmath}
\usepackage{amsfonts}
\usepackage{graphicx}
\usepackage{mathrsfs}
\usepackage{geometry}
\usepackage{doublespace}
\usepackage{pstcol,pst-fill,pst-grad}

\newtheorem{question}{Question}
\newtheorem{definition}{Definition}
\newtheorem{lemma}{Lemma}
\newtheorem{theorem}{Theorem}
\newtheorem{remark}{Remark}
\newtheorem{corollary}{Corollary}
\newtheorem{example}{Example}

\def\proof{\noindent{\bf Proof --~}}
\def\cqfd{\hfill$\square$}
\def\sgn{\mathrm{sign}}
\def\G{\mathscr{G}}
\def\GG{\mathcal{G}}
\def\N{\mathbb{N}}
\def\1n{1,\dots,n} 
\def\F{{\tilde F}} 
\def\f{\tilde f} 
\def\phi{\varphi}

\begin{document}

~\\~\\~\\
\begin{center}
\begin{LARGE}
Negative circuits and sustained oscillations\\ in asynchronous
automata networks
\end{LARGE}
~\\~\\
\begin{large}
Adrien Richard
\end{large}
~\\~\\
Laboratoire I3S, UMR 6070 CNRS \& Universit\'e de
Nice-Sophia Antipolis,\\[-4mm] 2000 route des Lucioles, 06903 Sophia
Antipolis, France.
~\\~\\
e-mail: \texttt{richard@i3s.unice.fr}\\[-4mm]
telephone: +33 4 92 94 27 51\\[-4mm]
fax: +33 4 92 94 28 98 
~\\~\\
\end{center}

\newpage

~\\~\\~\\

\paragraph{Abstract:}

The biologist Ren\'e Thomas conjectured, twenty years ago, that the
presence of a negative feedback circuit in the interaction graph of a
dynamical system is a necessary condition for this system to produce
sustained oscillations. In this paper, we state and prove this
conjecture for asynchronous automata networks, a class of discrete
dynamical systems extensively used to model the behaviors of gene
networks. As a corollary, we obtain the following fixed point theorem:
given a product $X$ of $n$ finite intervals of integers, and a map $F$
from $X$ to itself, if the interaction graph associated with $F$ has
no negative circuit, then $F$ has at least one fixed point.



\newpage

\section{Introduction}

We are interested in a class of discrete dynamical systems used to
model gene networks. The biological context is the following. Gene
networks are often described by Biologists under the form of
{\emph{interaction graphs}}. These are directed graphs where vertices
correspond to genes and where arcs are labelled with a sign: a
positive (negative) arc from a gene~$j$ to a gene~$i$ means that the
protein encoded by the gene~$j$ activates (represses) the synthesis of
the protein encoded by the gene~$i$. These very coarse descriptions of
gene networks are then taken as a basis to design much more complex
{\emph{dynamical models}} that describe the temporal evolution of the
concentration of the encoded proteins {\cite{dJ02}}. Unfortunately,
these models require, in most cases, unavailable informations on the
strength of the interactions. In this context, a difficult and
interesting question is: {\emph{which dynamical properties of a gene
network can be deduced from its interaction graph?}}

The biologist Ren\'e Thomas stated two well known conjectures that
partially answer this question. These conjectures can be informally
stated as follows {\cite{T81,KST07}}:
\begin{enumerate}
\item
The presence of a {\emph{positive circuit}} in the interaction graph
of a network ({\emph{i.e.}} a circuit with an {\emph{even}} number of
negative arcs) is a necessary condition for the presence of
{\emph{multiple stable states}} in the dynamics of the network.
\item
The presence of a {\emph{negative circuit}} in the interaction graph
of a network ({\emph{i.e.}} a circuit with an {\emph{odd}} number of negative
arcs) is a necessary condition for the presence of {\emph{sustained
oscillations}} in the dynamics of the network.
\end{enumerate}
It is worth noting that multistationarity and sustained oscillations
are, from a biological point of view, important dynamical properties often
related to differentiation processes and homeostasis phenomena
respectively {\cite{T81,TA90,TK01}}.

The first conjecture has been formally stated and proved by several
authors in continuous frameworks {\cite{PM95,G98,S98,CD02,S03,S06}},
in which the concentration of each protein evolves continuously,
generally following an ordinary differential equation system. The
first conjecture has been more recently stated and proved in discrete
frameworks {\cite{ADG04,A08,RC07,RR08,R09}}, in which the
concentration level of each protein evolves inside a
finite interval of integers, which is $\{0,1\}$ in the Boolean
case. Studies of the second conjecture are fewer: a Boolean version of
the second conjecture has been stated and proved by Remy, Ruet and
Thieffry {\cite{RR08}}, and there are only partial results in the
continuous case {\cite{G98,S98}}.

In this paper, we state and prove second Thomas' conjecture for
asynchronous automata networks (Theorem 1). Our interest for these
discrete dynamical systems comes from the fact that they have been
proposed by Thomas as model for the dynamics of gene networks more
than thirty years ago {\cite{T73,TA90,T91,TK01}}. They are still
extensively used because of the qualitative nature of most reliable
experimental data, and the fact that the sigmoidal shape of genetic
regulations leads to a natural discretization of concentrations
{\cite{GK73,S89,TA90,ST93,dJ04}}.

The discrete version of Thomas' conjecture we establish generalizes in
several ways the one established by Remy, Ruet and Thieffry
{\cite{RR08}} in the Boolean case: both the discrete dynamical
framework and the considered class of sustained oscillations are more
general. Furthermore, the class of sustained oscillations we consider
allows us to obtain, as an immediate consequence, the fixed point
theorem mentioned in the abstract (Corollary~1).

The paper is organized as follows. Section 2 presents definitions
related to asynchronous automata networks. In Section 3, second
Thomas' conjecture is stated and proved for these networks. In
Section~4, we establish a variant of second Thomas' conjecture 
more suited to the modeling of gene networks. Counter examples to
natural extension of the established results are given in Section 5.

\section{Definitions}

We consider a network of $n$ interacting automata, denoted from $1$ to
$n$. The set of possible states for automaton $i$ is a finite
intervals of integers $X_i$ of cardinality at least two. The set of
possible states for the network is the Cartesian product
$X=\prod_{i=1}^n X_i$. The dynamics of the network is then described
according to a map $F:X\to X$,
\[
x=(x_1,\dots,x_n)\in X~\mapsto~ F(x)=(f_1(x),\dots,f_n(x))\in X,
\]
with which we associate the maps $F_i:X\to X$ defined by
\[
F_i(x)=(x_1,\dots,x_{i-1},f_i(x),x_{i+1},\dots,x_n)\qquad (i=\1n).
\]
More precisely, given an initial point $x^0\in X$ and a map $\phi$
from $\N$ to $\{\1n\}$, the dynamics of the network is described by
the following recurrence, that we call the {\emph{asynchronous
iteration of $F$ induced by the strategy $\phi$ from initial point
$x^0$}}:
\begin{equation}\label{it}
x^{t+1}=F_{\phi(t)}(x^t)\qquad (t=0,1,2\dots). 
\end{equation}
Generally, one only considers the asynchronous iterations induced by
{\emph{pseudo-periodic strategies}}, {\emph{i.e.}} strategies $\phi$
such that $|\phi^{-1}(i)|=\infty$ for $i=\1n$ {\cite{R95,BM00}}.

In this paper, we will study the asynchronous iterations of $F$
through a directed graph on $X$ called the asynchronous state
transition graph of $F$. Before defining this graph, let us set, for
all $x\in X$,
\[
I_F(x)=\{i\in\{\1n\}~|~f_i(x)\neq x_i\}.
\]

\begin{definition}
The {\emph{asynchronous state transition graph of $F$}}, denoted
$\Gamma(F)$, is the directed graph whose set of vertices is $X$ and
whose set of arcs is
\[
\{(x,F_i(x))\,|\, x\in X,~i\in I_F(x)\}.
\]
\end{definition}

\begin{remark}
{\emph{$|I_F(x)|$ is the number of successors of $x$ in $\Gamma(F)$,
and $|I_F(x)|=0$ if and only if $x$ is a fixed point of $F$. Also,
$\Gamma(F)$ has no arc from a vertex to itself, and in the following,
we assume, by convention, that $\Gamma(F)$ has a path of length zero
from each vertex to itself.}}
\end{remark}
%
The relation between $\Gamma(F)$ and the asynchronous iterations of
$F$ is clear: there is a path from $x$ to $y$ in $\Gamma(F)$ if and
only if there exists a strategy $\phi$ such that the asynchronous
iteration of $F$ induced by $\phi$ from $x$ reaches $y$.

In this context, the fixed points of $F$ are of particular interest:
they correspond to the stable states of the system. More precisely, if
$\phi$ is a pseudo-periodic strategy, then the asynchronous iteration
(\ref{it}) stabilizes on a point $\xi$ (\emph{i.e.} there exists $t$
such that $x^{t}=x^{t+1}=\xi$) if {\emph{and only if}} $\xi$ is a
fixed point of $F$. In the following definition, we introduce a notion
of an attractor, which extends in a natural way the one of a stable
state.

\begin{definition}
A {\emph{trap domain}} of $\Gamma(F)$ is a non-empty subset
$D\subseteq X$ such that for every arc $(x,y)$ of $\Gamma(F)$, if
$x\in D$ then $y\in D$. An {\emph{attractor}} of $\Gamma(F)$ is a
smallest trap domain with respect to the inclusion. A {\emph{cyclic
attractor}} is an attractor of cardinality at~least~two.
\end{definition}

\begin{remark} 
{\emph{One has the three following basic properties: (1) $x$ is a
fixed point of $F$ if and only if $\{x\}$ is an attractor of
$\Gamma(F)$; (2) attractors perform an attraction in the weak sense
that, from any state, there always exists a path leading to one of
them; (3) if $x$ and $y$ belong to the same attractor, then there
exists a path from $x$ to $y$.}}
\end{remark}
%
The third point highlights the fact that inside a cyclic attractor,
each state has at least one successor. So, when the network is inside
a cyclic attractor, it cannot reach a fixed point, and thus, it
describes sustained oscillations. More precisely, {\emph{if $x^0$
belongs to a cyclic attractor $A$, then for all pseudo-periodic
strategy $\phi$, the asynchronous iteration of $F$ induced by $\phi$
from $x^0$ never leaves $A$ and never stabilizes, and since $A$ is
finite, it necessarily describes sustained oscillations}}. In the
following, we are interested in the relationships between sustained
oscillations produced by cyclic attractors and the negative circuits
of the interaction graph of the network.

An {\it interaction graph} is here defined to be a directed graph
whose set of vertices is $\{\1n\}$ and where each arc is provided with
a sign. Formally, each arc is characterized by a triple $(j,s,i)$
where $j$ ($i$) is the initial (final) vertex, and where
$s\in\{-1,1\}$ is the sign of the arc. An interaction graph can then
have both a positive and a negative arc from one vertex to another.

In the following definition, we attach to $F$ an interaction graph
$G(F)$ that is nothing but the interaction graph of the network whose
dynamics is described by the asynchronous iterations of $F$.

\begin{definition}
The {\emph{interaction graph of $F$}}, denoted $G(F)$, is the
interaction graph that contains a positive (negative) arc from $j$ to
$i$ if there exists $x \in X$ with $x_j+1\in X_j$~such~that
\[
f_i(x_1,\dots,x_j+1,\dots,x_n)-f_i(x_1,\dots,x_j,\dots,x_n)
\]
is positive (negative).
\end{definition}

\begin{remark}
{\emph{$G(F)$ has at least one arc from $j$ to $i$ if and only if
$f_i$ depends on $x_j$.}}
\end{remark}

\begin{definition}
A {\emph{path of $G(F)$}} of length $r\geq 1$ is a sequence of $r$
arcs of $G(F)$, say $(j_1,s_1,i_1),(j_2,s_2,i_2),\dots,(j_r,s_r,i_r)$,
such that $i_q=j_{q+1}$ for all $1\leq q<r$. Such a path is a path
{\emph{from $j_1$ to $i_r$}} of {\emph{sign}} $s=\prod_{q=1}^r
s_q$. It is a {\emph{circuit}} if $i_r=j_1$ and it is an
{\emph{elementary circuit}} if, in addition, the vertices $i_q$ are
mutually distinct.
\end{definition}

\begin{remark}
{\emph{If $G(F)$ has a negative circuit, then it has an elementary
negative circuit (this is false for positive circuits). So, in order
to prove that $G(F)$ has an elementary negative circuit, it is
sufficient to prove that $G(F)$ has a negative circuit.}}
\end{remark}

\begin{example}
$n=2$, $X=\{0,1,2\}^2$ and $F$ is defined by the following table:
\[
\begin{array}{c|ccccccccc}
x    &(0,0)&(0,1)&(0,2)&(1,0)&(1,1)&(1,2)&(2,0)&(2,1)&(2,2)\\\hline
F(x) &(2,0)&(1,0)&(0,2)&(2,0)&(0,0)&(0,1)&(2,1)&(0,1)&(0,1)
\end{array}
\]
The asynchronous state transition graph and the interaction graph of $F$~are~as~follows:
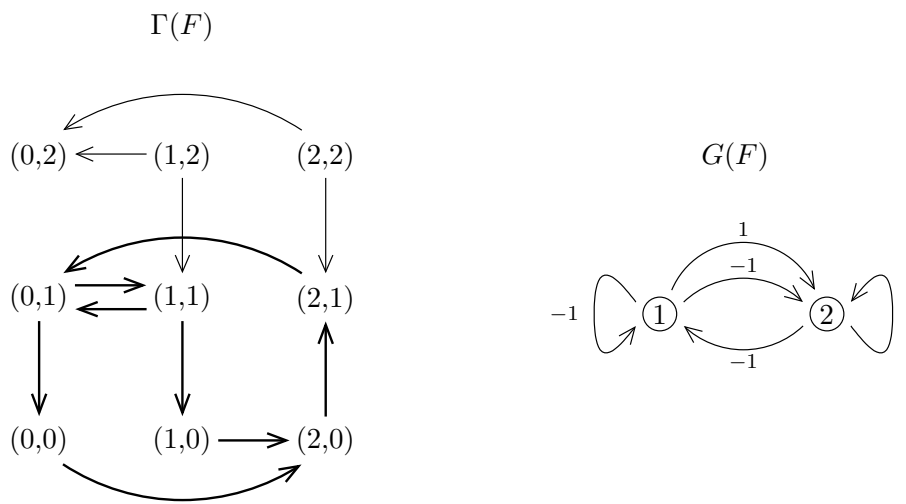
\[
\begin{array}{c}
\Gamma(F)\\[5mm]
\input{ExDef.pstex_t}
\end{array}
~~~~~~~~~~~~~~~~
\begin{array}{c}
G(F)\\[5mm]
\input{GraphDef.pstex_t}
\end{array}
\]
We see that $\Gamma(F)$ has two attractors: the stable state $(0,2)$
and the cyclic attractor $\{0,1,2\}\times\{0,1\}$. We also see that
$G(F)$ has two elementary positive circuits, and two elementary
negative circuits.
\end{example}

\section{Main result}

In this section, we prove the following discrete
version of second Thomas' conjecture:

\begin{theorem}
If $\Gamma(F)$ has a cyclic attractor, then $G(F)$ has a negative circuit.
\end{theorem}

\begin{remark}
{\emph{This theorem has been proved by Remy, Ruet and Thieffry
{\cite{RR08}} in the Boolean case ({\emph{i.e.}} when $X$ is the
$n$-dimensional hypercube $\{0,1\}^n$) and under the rather strong
hypothesis that $\Gamma(F)$ contains a {\emph{stable cycle}}, that is,
a cyclic attractor $A$ in which each state has a unique successor
({\emph{i.e.}} $|I_F(x)|=1$ for all $x\in A$).}}
\end{remark}

Before proving Theorem 1, let us point out that it has, as immediate
consequence, the following fixed point theorem (which can not be
deduced, in the Boolean case, from the theorem of Remy, Ruet and
Thieffry mentioned above):

\begin{corollary}
If $G(F)$ has a no negative circuit, then $F$ has at least one fixed
point.
\end{corollary}

\proof 
Indeed, if $F$ has no fixed point, then $\Gamma(F)$ has clearly at
least one cyclic attractor, and following Theorem 1, $G(F)$ has a
negative circuit.
\cqfd

\begin{remark}
{\emph{In {\cite[Chapter 13]{R95}} (see also {\cite{BM00}}), Robert
prove the following convergence result: if $G(F)$ has no circuit, then
$F$ has a unique fixed point $\xi$, and, for all initial point $x^0$
and for all pseudo-periodic strategy $\phi$, the asynchronous
iteration of $F$ induced by $\phi$ from $x^0$ reaches the fixed point
$\xi$. From Theorem 1 and the second point of Remark 2, one obtains a
convergence result that has a weaker conclusion under a weaker
condition: if $G(F)$ has no negative circuit, then $F$ has at least
one fixed point, and for all initial point $x^0$, there exists a
strategy $\phi$ for which the asynchronous iteration (\ref{it})
reaches a fixed point of $F$.}}
\end{remark}

The proof of Theorem 1 needs few additional definitions and
notations. Let $\GG$ and $\GG'$ be two interaction graphs with arc-set
$E$ and $E'$ respectively. We say that $\GG$ is a {\emph{subgraph}} of
$\GG'$ if $E\subseteq E'$. We denote by $\GG\cup\GG'$ the interaction
graph whose set of arcs is $E\cup E'$. Next, for all $x\in X$, we set
\[
f'_i(x)=\sgn(f_i(x)-x_i)\qquad (i=\1n),
\]
where $\sgn$ is the usual sign function ($\sgn(a)=a/|a|$ for all
$a\neq 0$, and $\sgn(0)=0$). The main tool used in the proof of
Theorem 1 is the following notion of local interaction graph:

\begin{definition}
For all $x\in X$, we denote by $\G_F(x)$ the interaction graph that
contains an arc from $j$ to $i$ of sign $s\in\{-1,1\}$ if
\[
f'_i(x)\neq f'_i(F_j(x))\quad\textrm{and}\quad s=f'_j(x)f'_i(F_j(x))
\]
\end{definition}

\begin{lemma}{\label{lem:subgraph}}
For all $x\in X$, $\G_F(x)$ is a subgraph of $G(F)$.
\end{lemma}

\proof
Let $x\in X$, and suppose that $\G_F(x)$ has an arc from $j$ to $i$ of
sign~$s$. For all integer $p$, we set
\[
x^p=(x_1,\dots,x_{j-1},x_j+p,x_{j+1},\dots,x_n).
\]

\noindent
{\bf Case \boldmath$j\neq i$\unboldmath.} By definition, $f'_j(x)\neq
0$. We suppose $f'_j(x)>0$ the other case being similar. Setting
$q=f_j(x)-x_j$, we have $q>0$ and $x^q=F_j(x)$. So $s=f'_i(x^q)$ and
$f'_i(x)=f'_i(x^0)\neq f'_i(x^q)$. Consider the smallest $0\leq p\leq
q$ such that $f'_i(x^p)=f'_i(x^q)$. Clearly, $p>0$ and
$f'_i(x^{p-1})\neq f'_i(x^p)=s$. So if $s=1$ then $f_i(x^{p-1})\leq
x_i<f_i(x^p)$ and we deduce that $G(F)$ has a positive arc from $j$ to
$i$. Similarly, if $s=-1$ then $f_i(x^{p-1})\geq x_i>f_i(x^p)$ and we
deduce that $G(F)$ has a negative arc from $j$ to $i$.\bigskip

\noindent
{\bf Case \boldmath$j= i$\unboldmath.} By definition,
$s=f'_i(x)f'_i(F_i(x))$ and $f'_i(x)\neq f'_i(F_i(x))$ thus
$s=-1$.~Suppose that $f'_i(x)>0$, the other case being similar. Then
$q= f_i(x)-x_i>0$ and $f'_i(F_i(x))<0$. Since $x^0=x$ and
$x^q=F_i(x)$, we deduce that $x^0_i<f_i(x^0)=x^q_i$ and
$f_i(x^q)<x^q_i$. Thus, there exists a smallest $0\leq p\leq q$ such
that $f_i(x^p)< x^q_i$. Clearly, $p>0$ and $x^{q}_i\leq
f_i(x^{p-1})$. Thus $f_i(x^p)<f_i(x^{p-1})$ and we deduce that $G(F)$
has a negative arc from $i$ to itself.
\cqfd

\begin{lemma}\label{lem:path}
Let $(x^0,x^1,\dots,x^r)$ be an elementary path of $\Gamma(F)$ of
length $r\geq 1$, and let $i\in I_F(x^r)$. If $f'_i(x^p)\neq
f'_i(x^r)$ for all $0\leq p<r$, then there exists $j\in
I_F(x^0)$~such~that $\bigcup_{q=0}^{r-1} \G_F(x^q)$ has a path from
$j$ to $i$ of sign $f'_j(x^0)f'_i(x^r)$.
\end{lemma}

\proof 
We reason by induction on the length $r$ of the path.\bigskip

\noindent
{\bf Case \boldmath$r=1$\unboldmath}. Since $(x^0,x^1)$ is an arc of
$\Gamma(F)$ there exists $j\in I_F(x^0)$ such that
$x^1=F_j(x^0)$. Following the conditions of the lemma $f'_i(x^0)\neq
f'_i(x^1)$, and thus, by definition, $\G_F(x^0)$ has an arc from $j$
to $i$ of sign $f'_j(x^0)f'_i(x^1)$.\bigskip
	
\noindent
{\bf Case \boldmath$r>1$\unboldmath}. Since $(x^{r-1},x^r)$ is a path
of $\Gamma(F)$ of length $1$ satisfying the conditions of the lemma
for $i\in I_F(x^r)$, following the base case, there exists $k\in
I_F(x^{r-1})$ such that $\G_F(x^{r-1})$ has a path from $k$ to $i$ of
sign
\[
s_{ki}=f'_k(x^{r-1})f'_i(x^r).
\] 
Now, consider the smallest $0\leq p<r$ such that
$f'_k(x^p)=f'_k(x^{r-1})$. First, suppose that $p=0$. Then $k\in
I_F(x^0)$ and $f'_k(x^0)f'_i(x^r)$ is equals to sign $s_{ki}$ of the
path of $\G_F(x^{r-1})$ from $k$ to $i$ mentioned above, so that the
lemma holds. Now, suppose that $p>0$. Then, by the choice of $p$, for
all $0\leq l< p$, we have $f'_k(x^l)\neq f'_k(x^p)$. Thus, the path
$(x^0,\dots,x^p)$ satisfies the conditions of the lemma for $k\in
I_F(x^p)$. Since $p<r$, by induction hypothesis, there exists $j\in
I_F(x^0)$ such that $\bigcup_{q=0}^{p-1} \G_F(x^q)$ has a path from
$j$ to $k$ of sign
\[
s_{jk}=f'_j(x^0)f'_k(x^p).
\]
Since $\G_F(x^{r-1})$ contains a path from $k$ to $i$ of sign
$s_{ki}$, we deduce that $\bigcup_{q=0}^{r-1} \G_F(x^q)$ contains a
path from $j$ to $i$ of sign
\[
s_{ji}=s_{jk}s_{ki}=f'_j(x^0)f'_k(x^p)f'_k(x^{r-1})f'_i(x^r), 
\]
and since $f'_k(x^p)=f'_k(x^{r-1})$, we deduce that
$s_{ji}=f'_j(x^0)f'_i(x^r)$.
\cqfd

\begin{lemma}\label{lem:base}
Let $A$ be a cyclic attractor of $\Gamma(F)$. If there exists $x\in A$
such that $|I_F(x)|=1$ then $\bigcup_{x\in A} \G_F(x)$ has a negative
circuit.
\end{lemma}

\proof
Suppose that there exists $x^0\in A$ such that $|I_F(x^0)|=1$, and let
$i$ be the unique element of $I_F(x^0)$. Suppose that $f'_i(x^0)>0$,
the other case being similar. Let $x^1=F_i(x)$. Then $\Gamma(F)$ has
an arc from $x^0$ to $x^1$ and we have $x^0_i<x^1_i$. Since $x^0\in
A$, we have $x^1\in A$, and we deduce that $\Gamma(F)$ has an
elementary path $(x^1,x^2,\dots,x^r)$ from $x^1$ to $x^r=x^0$ whose
all the vertices belong to $A$. If $f'_i(x^p)\geq 0$ for all $0<p<r$,
then $x^p_i\leq x^{p+1}_i$ for all $0<p<r$, and we deduce that
$x^1_i\leq x^r_i=x^0_i$, a contradiction. Thus, there exists a
smallest $0<p< r$ such that $f'_i(x^p)<0$. Then, $(x^0,x^1,\dots,x^p)$
is an elementary path where $i\in I_F(x^p)$ and by the choice of $p$,
we have $f'_i(x^l)\neq f'_i(x^p)$ for all $0\leq l<p$. So, according
to Lemma~{\ref{lem:path}}, there exists $j\in I_F(x^0)$ such that
$\bigcup_{q=0}^{p-1}\G_F(x^q)$ contains a path from $j$ to $i$ of sign
$f'_j(x^0)f'_i(x^p)$. Since $I_F(x^0)=\{i\}$, we have $j=i$ and
consequently, $\bigcup_{q=0}^{p-1}\G_F(x^q)$ contains a path from $i$
to itself, and thus a circuit, of sign $f'_i(x^0)f'_i(x^p)$. By
construction, $f'_i(x^0)f'_i(x^p)<0$, thus this circuit is negative,
and since $\{x^0,\dots,x^{p-1}\}\subseteq A$, it is contained in
$\bigcup_{x\in A}\G_F(x)$.\cqfd

\begin{lemma}\label{lem:induction}
Let $A$ be a cyclic attractor of $\Gamma(F)$. If $|I_F(x)|>1$ for all
$x\in A$, then there exists $H:X\to X$ such that $\Gamma(H)$ contains
a cyclic attractor strictly included in $A$, and such that $\G_H(x)$
is a subgraph of $\G_F(x)$ for all $x\in X$.
\end{lemma}

\proof 
Suppose $A$ to be a cyclic attractor of $\Gamma(F)$ such that
$|I_F(x)|>1$ for all $x\in A$. Let $y$ be any state of $A$. Then
$I_F(y)$ contains at least two elements, and without loss of
generality, we can suppose that $1\in I_F(y)$. Consider the map
$H:X\to X$ defined by:
\[
\forall x\in X,\qquad 
H(x)=(h_1(x),h_2(x),\dots,h_n(x))=
(x_1,f_2(x),\dots,f_n(x)).
\]

We first prove that $A$ is a trap domain of $\Gamma(H)$. For that, it
is sufficient to prove that, given any $x\in A$ and $i\in I_H(x)$, we
have $H_i(x)\in A$. Since $h_1(x)=x_1$, $1\not\in I_H(x)$, so $i\neq
1$. Thus $F_i(x)=H_i(x)$, and since $A$ is a trap domain of
$\Gamma(F)$, we have $F_i(x)\in A$ and we deduce that $H_i(x)\in A$ as
expected. So $A$ is a trap domain of $\Gamma(H)$ and, by definition,
$\Gamma(H)$ contains at least one attractor $B\subseteq A$.

We claim that $B$ is a cyclic attractor of $\Gamma(H)$. Let $x\in
B$. Then $x\in A$ so $|I_F(x)|>1$ and we deduce that $I_F(x)$ contains
an index $i\neq 1$. Then, $x_i\neq f_i(x)=h_i(x)$ so $x\neq
H_i(x)$. Since $x\in B$ we have $H_i(x)\in B$. So $|B|\geq 2$,
{\emph{i.e.}} $B$ is a cyclic attractor of $\Gamma(H)$.

We now prove that $B\subset A$ (strict inclusion). Suppose, by
contradiction, that $B=A$. Since $1\in I_F(y)$ and $y\in A$, we have
$y\neq F_1(y)\in A=B$. Since $B$ is an attractor of $\Gamma(H)$, we
deduce that $\Gamma(H)$ has a path $(x^0,x^1,\dots,x^r)$ from $x^0=y$
to $x^r=F_1(y)$. Since $h_1(x)=x_1$ for all $x\in X$, we have
$x^0_1=x^1_1=\dots=x^r_1$. So $y_1=f_1(y)$, a contradiction.

It remains to prove that $\G_H(x)$ is a subgraph of $\G_F(x)$ for all
$x\in X$.  If $(j,s,i)$ is an arc of $\G_H(x)$, then by definition,
$h'_j(x)\neq 0$ and $h'_i(H_j(x))\neq 0$. So $j\neq 1$ and $i\neq
1$. Thus $f_j=h_j$ and $f_i=h_i$. It is then clear that $(i,s,j)$ is
an arc of $G_F(x)$.\cqfd

\begin{lemma}\label{lem:main}
If $A$ is a cyclic attractor of $\Gamma(F)$, then $\bigcup_{x\in
A}\G_F(x)$ has a negative circuit.
\end{lemma}

\proof
Let $U$ be the set of couples $(F,A)$ such that $F$ is a map from $X$
to itself, and such that $A$ is a cyclic attractor of $\Gamma(F)$. Let
$\prec$ be the well funded strict order on $U$ defined by
$(H,B)\prec(F,A)$ if and only if $B$ is strictly included in
$A$. Reasoning by induction on the set $U$ ordered by $\prec$, we show
that, for all $(F,A)\in U$, $\bigcup_{x\in A} \G_F(x)$ has a negative
circuit.\bigskip

\noindent
{\bf Base case.} Let $(F,A)$ be a minimal element of $(U,\prec)$. If
$|I_F(x)|>1$ for all $x\in A$, then, following Lemma
{\ref{lem:induction}}, there exists $(H,B)\in U$ such that $(H,B)\prec
(F,A)$, and this contradict the minimality of $(F,A)$. So there exists
$x\in A$ such that $|I_F(x)|=1$ and, following Lemma~{\ref{lem:base}},
$\bigcup_{x\in A} \G_F(x)$ has a negative circuit.\bigskip

\noindent
{\bf Induction step.} Let $(F,A)$ be a non-minimal element of
$(U,\prec)$. By induction hypothesis, for all $(H,B)\prec (F,A)$,
$\bigcup_{x\in B} \G_H(x)$ has a negative circuit. If, for all $x\in
A$, we have $|I_F(x)|> 1$, then following Lemma {\ref{lem:induction}},
there exists $(H,B)\prec (F,A)$ such that $\G_H(x)$ is a subgraph of
$\G_F(x)$ for all $x\in X$. Since $B\subset A$, we deduce that
$\bigcup_{x\in B}\G_H(x)$ is a subgraph of $\bigcup_{x\in A} \G_F(x)$,
and since, by induction hypothesis, $\bigcup_{x\in B}\G_H(x)$ has a
negative circuit, we deduce that $\bigcup_{x\in A} \G_F(x)$ has a
negative circuit. Otherwise, there exists $x\in A$ such that
$|I_F(x)|=1$, and following Lemma {\ref{lem:base}}, $\bigcup_{x\in A}
\G_F(x)$ has again a negative circuit.~\cqfd\bigskip

\noindent
{\bf Proof of Theorem 1 --~} If $A$ is a cyclic attractor of
$\Gamma(F)$, then by Lemma~{\ref{lem:main}}, $\bigcup_{x\in A}\G_F(x)$
has a negative circuit. By Lemma~{\ref{lem:subgraph}}, $\bigcup_{x\in
A}\G_F(x)$ is a subgraph of $G(F)$ and we deduce that $G(F)$ has a
negative circuit.\cqfd

\begin{remark}
{\emph{The key lemma is clearly Lemma~5, which shows that it is
sufficient to consider the restriction of $F$ to a cyclic attractor
$A$ in order to obtain a negative circuit.}}
\end{remark}

\section{A variant for gene regulatory networks}

In this section, we establish a variant of Theorem 1 that is more
suited to the modeling of gene networks. To model the behaviors of a
network of $n$ genes, Thomas {\cite{T73,TA90,TK01}} proposes to
consider an ``unitary'' asynchronous state transition graph
$\Gamma[F]$ that is slightly different than $\Gamma(F)$. In
$\Gamma[F]$, each transition starting from a given state $x$ involves,
as in $\Gamma(F)$, the evolution of the state $x_i$ of at most one
component $i\in I_F(x)$, but in $\Gamma[F]$, this state $x_i$ is not
updated to $f_i(x)$: it is increased or decreased by a unit depending
on whether $x_i<f_i(x)$ or $x_i>f_i(x)$. Thanks to this updating rule,
unitary asynchronous state transition graphs can be seen as
discretizations of piece-wise linear differential equation systems
{\cite{S89,ST93}}.

\begin{definition}
The {\emph{unitary}} asynchronous state transition graph of $F$,
denoted $\Gamma[F]$, is the asynchronous state transition graph
$\Gamma(\F)$ of the map $\F:X\to X$ defined~by
\[
\F(x)=(\f_1(x),\dots,\f_n(x)),\qquad \f_i(x)=x_i+f'_i(x)\qquad (i=\1n).
\]
\end{definition}

\begin{remark}
{\emph{In the Boolean case, $\Gamma[F]=\Gamma(F)$.}}
\end{remark}

We are now confronted to the following problem: $G(F)$ cannot be seen
as {\emph{the}} interaction graph of the network whose dynamics is
described by $\Gamma[F]$, since maps $H$ such that $G(H)\neq G(F)$ and
$\Gamma[H]=\Gamma[F]$ may exist. In addition, it is not satisfactory
to see $G(\F)$ as the interaction graph of the network whose dynamics
is described by $\Gamma[F]$, since maps $H$ such that $G(H)$ is a
{\emph{strict}} subgraph $G(\F)$ and such that $\Gamma[H]=\Gamma[F]$
may also exist.

To solve this problem, Richard and Comet {\cite{RC07}} define a
subgraph $G[F]$ of $G(F)$ that only depends on $\Gamma[F]$ and
provide, in this way, a natural and non-ambiguous definition of the
interaction graph of the network whose dynamics is described by
$\Gamma[F]$. Furthermore, one can show that $G[F]$ is, with respect to
the subgraph relation, the smallest interaction graph from which one
can obtain $\Gamma[F]$ by following the logical method developed by
Thomas to model gene networks {\cite{R06}}.

\begin{definition}
We denote by $G[F]$ the interaction graph that contains a positive arc
from $j$ to $i$ if there exists $x\in X$ with $x_j+1\in X_j$ such that
\[
f_i(x_1,\dots,x_j,\dots,x_n)\leq x_i <f_i(x_1,\dots,x_j+1,\dots,x_n),
\]
and that contains a negative arc from $j$ to $i$ if there exists $x\in X$ with
$x_j+1\in X_j$ such~that
\[
f_i(x_1,\dots,x_j,\dots,x_n)> x_i \geq f_i(x_1,\dots,x_j+1,\dots,x_n).
\]
\end{definition}

\begin{remark}
{\emph{$G[F]$ is a subgraph of $G(F)$, and in the Boolean case,
$G[F]=G(F)$.}}
\end{remark}

We now establish, in this setting, the following discrete version of
second Thomas' conjecture (which is, as Theorem 1, an immediate
consequence of Lemma~{\ref{lem:main}}):

\begin{theorem}
If $\Gamma[F]$ has a cyclic attractor, then $G[F]$ has a negative
circuit.
\end{theorem}

\begin{lemma}
For all $x\in X$, $\G_{\F}(x)$ is a subgraph of $G[F]$.
\end{lemma}

\proof
First observe that $f'_i(x)=\f'_i(x)$ for all $x\in X$ and
$i\in\{\1n\}$. Furthermore, if $\f_i(x)\leq x_i$ (resp. $\f_i(x)\geq
x_i$) then $f_i(x)\leq\f_i(x)$ (resp. $f_i(x)\geq
\f_i(x)$).

Now, suppose that $\G_\F(x)$ has an arc from $j$ to $i$ of sign $s$
with $j\neq i$. Let
\[
y=(x_1,\dots,x_j+\f'_j(x),\dots,x_n)
\]
and observe that $y=\F_j(x)$. Suppose that $\f'_i(y)>0$, the other
case being similar. Then, by definition, $\f'_j(x)=s$ and
$\f'_i(x)\leq 0$. Thus $\f_i(x)\leq x_i=y_i < \f_i(y)$ and we deduce
that
\[
f_i(x)\leq \f_i(x)\leq x_i=y_i < \f_i(y)\leq f_i(y).
\]
So if $\f'_j(x)=s$ is positive then
\[
f_i(x)\leq x_i<f_i(y)=f_i(x_1,\dots,x_j+1,\dots,x_n)
\]
and we deduce that $G[F]$ has a positive arc from $j$ to $i$, and if
$\f'_j(x)=s$ is negative then
\[
f_i(y_1,\dots,y_j+1,\dots,y_n)=f_i(x)\leq y_i<f_i(y)
\]
and we deduce that $G[F]$ has a negative edge from $j$ to $i$. 

Suppose now that $\G_\F(x)$ has an arc from $i$ to itself of sign
$s$. By definition, we have $s=\f'_i(x)\f'_i(\F_i(x))$ and
$\f'_i(x)\neq\f'_i(\F_i(x))$ so that $s$ is negative. Suppose that
$\f'_i(x)>0$, the other case being similar. Then,
$\F_i(x)=(x_1,\dots,x_i+1,\dots,x_n)$ and $\f'_i(\F_i(x))<0$. Thus
\[
\f_i(x_1,\dots,x_i+1,\dots,x_n)\leq x_i<\f_i(x)
\]
and we deduce that
\[
f_i(x_1,\dots,x_i+1,\dots,x_n)\leq 
\f_i(x_1,\dots,x_i+1,\dots,x_n)\leq x_i<\f_i(x)\leq f_i(x).
\] 
Consequently, $G[F]$ has a negative arc from $i$ to itself.
\cqfd\bigskip

\noindent
{\bf Proof of Theorem 2 --~} Since $\Gamma[F]=\Gamma(\F)$, if
$\Gamma[F]$ has a cyclic attractor $A$, then by Lemma 5,
$\bigcup_{x\in X}\G_\F(x)$ has a negative circuit. Following the
previous lemma, $\bigcup_{x\in X}\G_\F(x)$ is a subgraph of $G[F]$,
and we deduce that $G[F]$ has a negative circuit.\cqfd

\begin{corollary}
If $G[F]$ has a no negative circuit, then $F$ has at least one fixed
point.
\end{corollary}

\proof 
If $F$ has no fixed point, then $\Gamma[F]$ has at least one cyclic
attractor, and following Theorem 2, $G[F]$ has a negative circuit.
\cqfd

\begin{remark}
{\emph{Since $G[F]$ is a subgraph of $G(F)$, Corollary 2 is
stronger than Corollary~1 (the same conclusion is obtained under a
weaker condition). In addition, from Theorems~1 and 2, it is clear
that: {\emph{if $\Gamma(F)$ or $\Gamma[F]$ has a cyclic attractor, then
$G(F)$ has a negative circuit}}. This generalizes Theorem 1 (the same
conclusion is obtained under a weaker condition). Indeed, as showed by
the following two examples, the presence of a cyclic attractor in
$\Gamma(F)$ ($\Gamma[F]$) does not imply the presence of a
cyclic attractor in $\Gamma[F]$ ($\Gamma(F)$).}}
\end{remark}

\begin{example}
$n=1$, $X=\{0,1,2\}$ and $F$ defined by $F(0)=2$, $F(1)=1$ and
$F(2)=0$. The state transitions graphs $\Gamma(F)$ and $\Gamma(F)$ are
the following:
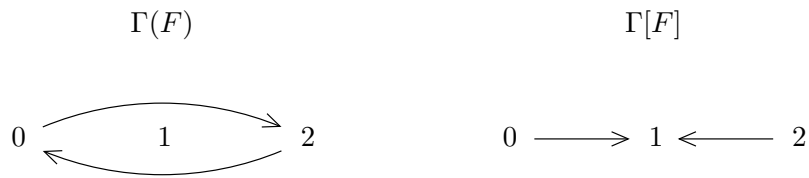
\[
\begin{array}{ccc}
\Gamma(F)&&\Gamma[F]\\[5mm]
\input{Ex_13.pstex_t}&~~~~~~~~~~&\input{Ex_14.pstex_t}
\end{array}
\]
We see that $\Gamma(F)$ has a cyclic attractor and that $\Gamma[F]$
has no cyclic attractor. The interaction graph $G(F)$ is the
interaction graph with one vertex and a negative arc from this vertex
to itself: it has thus a negative circuit. The interaction graph
$G[F]$ is the interaction graph with one vertex and no arc (it is a
strict subgraph of $G(F)$). This shows that the presence of a cyclic
attractor in $\Gamma(F)$ does not imply the presence of a negative
circuit~in~$G[F]$.
\end{example}

\begin{example}
$n=1$, $X=\{0,1,2\}$ and $F$ defined by $F(0)=0$, $F(1)=2$ and
$F(2)=0$. The state transitions graphs $\Gamma(F)$ and $\Gamma(F)$ are
the following:
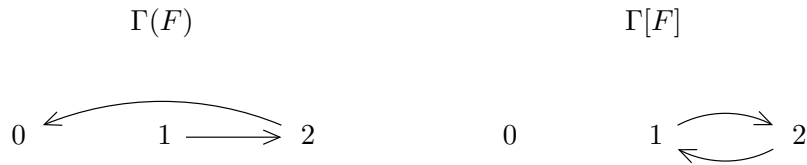
\[
\begin{array}{ccc}
\Gamma(F)&&\Gamma[F]\\[5mm]
\input{Ex_11.pstex_t}&~~~~~~~~~~&\input{Ex_12.pstex_t}
\end{array}
\]
We see that $\Gamma[F]$ has a cyclic attractor and that $\Gamma(F)$
has no cyclic attractor. The interaction graphs $G(F)$ and $G[F]$ are
equal to the interaction graph with one vertex and both a positive and
a negative arc from this vertex to itself ($G(F)$ and $G[F]$ have thus
a negative circuit).
\end{example}

{\section{Concluding remarks}}

The weakest condition allowing the asynchronous iterations of $F$ to
describe sustained oscillations is the presence of a directed cycle in
$\Gamma(F)$. However, as showed by the following example, the presence
of a directed cycle in $\Gamma(F)$ does not imply the presence of a
negative circuit in $G(F)$ (one can only show that it implies the
presence of a circuit in $G(F)$). This shows that structures in
$\Gamma(F)$ stronger than directed cycles (such as cyclic attractors)
are needed to obtain a negative circuit.

\begin{example} 
$n=3$, $X=\{0,1\}^3$ and $F$ is defined by 
\[
\begin{array}{l}
f_1(x)=x_3\\
f_2(x)=x_1\\
f_3(x)=x_2.
\end{array}
\]
The asynchronous state transition graph $\Gamma(F)$ (which is here
equal to $\Gamma[F]$) and the interaction graph $G(F)$ (which is here
equal to $G[F]$) are the following:
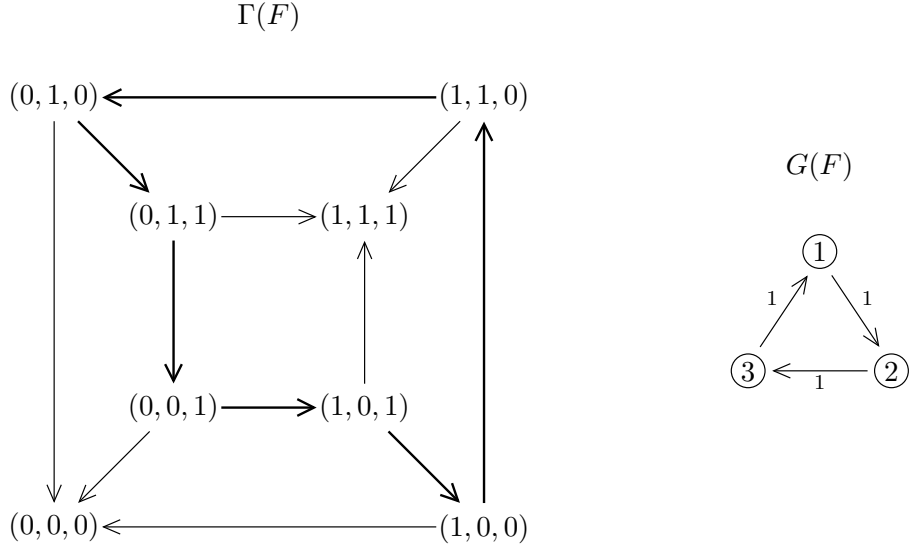
\[
\begin{array}{c}
\Gamma(F)\\[5mm]
\input{Ex1.pstex_t}
\end{array}
~~~~~~~~~~~~~~~~
\begin{array}{c}
G(F)\\[5mm]
\input{Graph1.pstex_t}
\end{array}
\]
We see that $\Gamma(F)$ has a directed cycle and that $G(F)$ has no
negative circuit.
\end{example}

A second remark is that it is not easy to find other classes of
iterations for which Theorem~1 remains valid. Consider for instance
the {\emph{synchronous state transition graph}} $\Lambda(F)$ that
encodes the behaviors of the iteration $x^{t+1}=F(x^t)$: the set of
vertices of $\Lambda(F)$ is $X$ and the set of its arcs is
$\{(x,F(x))~|~x\in X,x\neq F(x)\}$. The cyclic attractors of such a
(deterministic) state transition graph $\Lambda(F)$ are naturally
defined to be the directed cycles of $\Lambda(F)$. However, the
following example shows that the presence of a directed cycle in
$\Lambda(F)$ does not imply the presence of a negative circuit
in~$G(F)$ (Robert {\cite{R86,R95}} prove that it only implies the
presence of a circuit in $G(F)$).

\begin{example} 
$n=2$, $X=\{0,1\}^2$ and $F$ is defined by 
\[
\begin{array}{l}
f_1(x)=x_2\\
f_2(x)=x_1.
\end{array}
\]
The synchronous state transition graph $\Lambda(F)$ and the
interaction graph $G(F)$ are as follows:
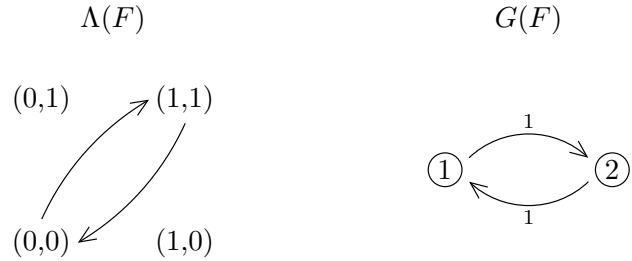
\[
\begin{array}{ccc}
\Lambda(F)&&G(F)\\[5mm]
\begin{array}{c}
\input{Ex2.pstex_t}
\end{array}
&~~~~~~~~~~~~~&
\begin{array}{c}
\input{Graph2.pstex_t}
\end{array}
\end{array}
\]
We see that $\Lambda(F)$ has a cyclic attractor and that $G(F)$ has no
negative circuit.
\end{example}

Finally, we can ask if, under the condition that $\Gamma(F)$ has a
cyclic attractor, a conclusion stronger than ``$G(F)$ has a negative
circuit'' could be obtained. Following Example 2, the presence of a
cyclic attractor in $\Gamma(F)$ does not imply the presence of a
negative circuit in the subgraph $G[F]$ of $G(F)$. So, another
direction has to be taken. As showed below, previous results on
the links between the interaction graph and the dynamical properties of
automata networks suggest to improve the conclusion of Theorem 1 by
studying if the presence of a cyclic attractor in $\Gamma(F)$ implies
the presence of a negative circuit in a {\emph{local interaction
graph}} associated with $F$.

\begin{definition}
For all $x\in X$, the {\emph{local interaction graph of $F$ evaluated
at state $x$}} is the interaction graph $G_F(x)$ that contains a
positive (negative) arc from $j$ to $i$ if $x_j+1\in X_j$~and
\[
f_i(x_1,\dots,x_j+1,\dots,x_n)-f_i(x_1,\dots,x_j,\dots,x_n)
\]
is positive (negative), or if $x_j-1\in X_j$ and 
\[
f_i(x_1,\dots,x_j,\dots,x_n)-f_i(x_1,\dots,x_j-1,\dots,x_n)
\]
is positive (negative).
\end{definition}

\begin{remark}
$G_F(x)$ is a subgraph of $G(F)$. More precisely, $G(F)=\bigcup_{x\in
X} G_F(x)$.
\end{remark}
%
With this material, Richard and Comet {\cite{RC07}} prove the
following local version of first Thomas' conjecture:

\begin{theorem}{\bf{\cite{RC07}}}~
If $\Gamma[F]$ has several attractors, and in particular if $F$ has
several fixed points, then there exists $x\in X$ such that $G_F(x)$
has a positive circuit.
\end{theorem}
%
Let us also mention the following fixed point theorem proved by
Richard {\cite{R08}} (and previously proved by Shih and Dong
{\cite{SD05}} in the Boolean case):

\begin{theorem}{\bf{\cite{R08}}}~
If $G_F(x)$ has no circuit for all $x\in X$, then $F$ has a unique
fixed point.
\end{theorem}
%
The proof of Theorem 4 done in {\cite{R08}} reveals that if $G_F(x)$
has no circuit for all $x\in X$, then $F$ has a unique fixed point
$\xi$, and, in addition, for all $x\in X$, $\Gamma[F]$ has a path from
$x$ to $\xi$. It is then clear that the presence of a cyclic attractor
in $\Gamma[F]$ implies the presence of a circuit in $G_F(x)$ for at
least one $x\in X$. We then arrive to the following natural question:

\begin{question}
Does the presence of a cyclic attractor in $\Gamma[F]$ or $\Gamma(F)$
implies the presence of a negative circuit in $G_F(x)$ for at least
one $x\in X$?
\end{question}
%
Clearly, a positive answer would improve significantly Theorem 1 or 2
by providing a local version of second Thomas' conjecture. However,
the following example shows that the answer is negative. This
highlights the fact that it is necessary to take a union of local
interaction graphs in order to obtain, from a cyclic attractor, a
negative circuit.

\begin{example} $n=2$, $X=\{0,1,2,3\}^2$ and $F$ is defined by:
\[
\begin{array}{l}
f_1(x)=
\left\{
\begin{array}{l}
3\textrm{ if $x_2=3$ or if $x_2>0$ and $x_1\geq 2$}\\
0\textrm{ otherwise} 
\end{array}
\right.
\\[13mm]
f_2(x)=
\left\{
\begin{array}{l}
3\textrm{ if $x_1=0$ or if $x_1<3$ and $x_2\geq 2$}\\
0\textrm{ otherwise} 
\end{array}
\right.
\end{array}
\]
The asynchronous state transition graph $\Gamma(F)$ is the following:
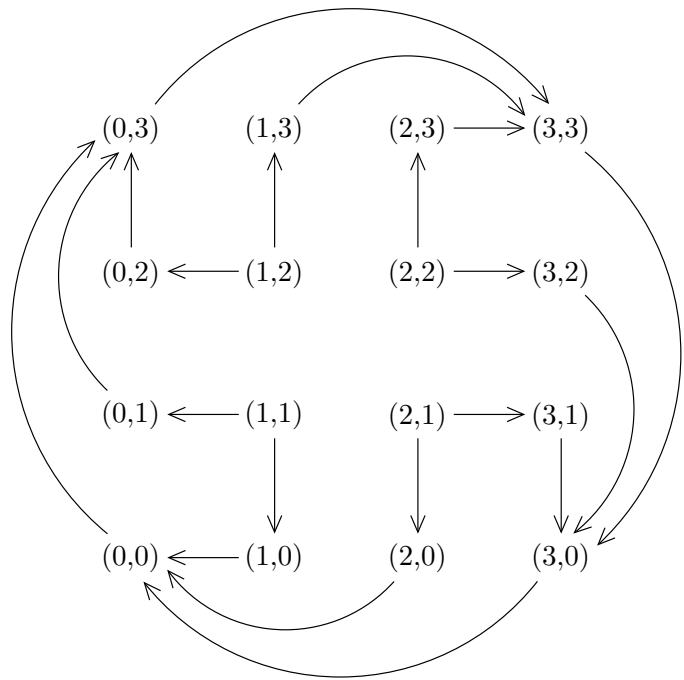
\[
\input{Ex3.pstex_t}
\]
The unitary asynchronous state transition graph $\Gamma[F]$ is the
following:
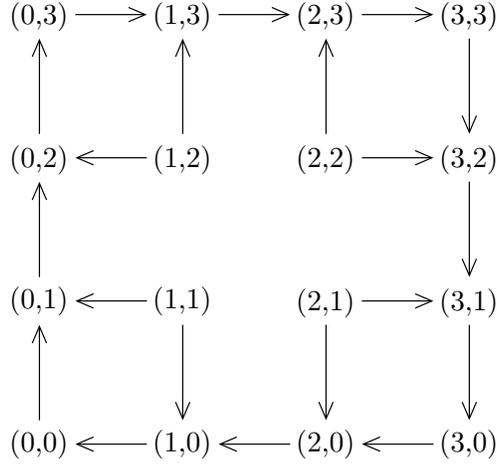
\[
\input{Ex3Bis.pstex_t}
\]
The interaction graph $G(F)$, which is here equal to $G[F]$, is the following:
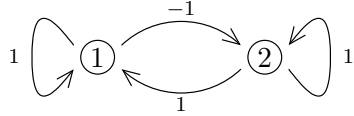
\[
\input{Graph3.pstex_t}
\]
We see that $\{(0,0),(0,3),(3,3),(3,0)\}$ is a cyclic attractor of
$\Gamma(F)$ and that $G(F)$ has a negative circuit. We see also that
$\{(0,0),(0,1),(0,2),(0,3),(1,3),(2,3),(3,3),(3,2),(3,1)(3,0)\}$ is a
cyclic attractor of $\Gamma[F]$ and that $G[F]$ has a negative
circuit. However, for all $x\in X$, the local interaction graph
$G_F(x)$ has no negative circuit. Indeed, for
$x\in\{(1,0),(0,0),(0,1)\}$ and $x\in\{(2,3),(3,3),(3,2)\}$, $G_F(x)$
is as follows:
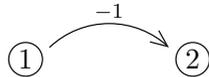
\[
\input{Graph3_00.pstex_t}
\]
for $x\in\{(3,1),(3,0),(2,0)\}$ and $x\in\{(0,2),(0,3),(1,3)\}$,
$G_F(x)$ is as follows:
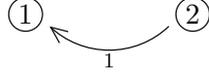
\[
\input{Graph3_30.pstex_t}
\]
for $x=(1,1)$ and $x=(2,2)$, $G_F(x)$ is as follows:
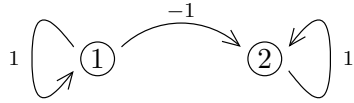
\[
\input{Graph3_11.pstex_t}
\]
and for $x=(1,2)$ and $x=(2,1)$, $G_F(x)$ is as follows:
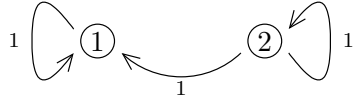
\[
\input{Graph3_12.pstex_t}
\]
\end{example}
%
The fact that Theorem 3 establishes the uniqueness of a fixed point for
$F$ under the condition that $G_F(x)$ has no positive circuit suggests
the following weaker version of Question~1:

\begin{question}
Does the absence of a negative circuit in $G_F(x)$ for all $x\in X$
implies the presence of at least one fixed point for $F$?
\end{question}
%
A positive answer would improve significantly Corollary 1, and would
give, together with Theorem 3, a very nice ``dichotomous'' proof of
Theorem 4. However, the previous example shows that Question 2 as also
a negative answer. Nevertheless,
\[
\textrm{{\emph{Questions 1 and 2 remain open in
the Boolean case}}}.
\]

{\subsection*{Acknowledgments}}

\noindent
Example 6 has been obtained with Jean-Paul Comet that I gratefully
thank. I also thank Bruno Soubeyran for stimulating discussions.




\begin{thebibliography}{7}

\bibitem{ADG04}%
J. Aracena, J. Demongeot, E. Goles, Positive and negative circuits in
discrete neural networks, {\emph{IEEE Trans. Neural
Networks}}, 15 (2004) 77-83.

\bibitem{A08}%
J. Aracena, Maximum number of fixed points in regulatory boolean
networks, {\emph{Bull. Math. Biol.}}, 70 (2008)
1398-1409.

\bibitem{BM00}
J. Bahi, C. Michel, Convergence of discrete asynchronous iterations,
{\emph{Int. J. Comput. Math.}}, 74 (2000) 113-125.

\bibitem{CD02}%
O. Cinquin, J. Demongeot, Positive and negative feedback: a striking
balance between necessary antagonists, {\emph{J. Theor. Biol.}}, 216
(2002) 229-241.

\bibitem{GK73}%
L. Glass, S.A. Kauffman, The logical analysis of continuous non linear
biochemical control networks, {\emph{J. Theor. Biol.}},
39 (1973) 103-129.

\bibitem{G98}%
J. L. Gouz\'e, Positive and negative circuits in dynamical systems,
{\emph{J. Biol. Syst.}}, 6 (1998) 11-15.

\bibitem{dJ02}%
H. de Jong, Modeling and simulation of genetic regulatory systems: a
literature review, {\emph{J. Comput. Biol.}}, 9 (2002)
67-103.

\bibitem{dJ04}%
H. de Jong, J.-L. Gouz\'e, C. Hernandez, M. Page, S. Tewfik,
J. Geiselmann, Qualitative simulation of genetic regulatory networks
using piecewise-linear models, {\emph{Bull. Math. Biol.}}, 66 (2004) 301-340.

\bibitem{KST07}%
M. Kaufman, C. Soul\'e, R. Thomas, A new necessary condition on
interaction graphs for multistationarity, {\emph{J.
Theor. Biol.}}, 248 (2007) 675-685.

\bibitem{PM95}%
E.~Plathe,~T.~Mestl,~S.W.~Omholt,~Feedback~loops,~stability~and
multistationarity in dynamical systems, {\emph{J. Biol. Syst.}}, 3
(1995) 569-577.

\bibitem{RR08}%
E. Remy, P. Ruet, D. Thieffry, Graphics requirement for multistability
and attractive cycles in a boolean dynamical framework,
{\emph{Adv. Appl. Math.}}, 41 (2008) 335-350.

\bibitem{R06}%
A. Richard, {\emph{Mod\`ele formel pour les r\'eseaux de r\'egulation
g\'en\'etique et influence des circuits de r\'etroaction}},
Ph.D. Thesis, University of Evry Val d'Essonne, France, 2006.

\bibitem{RC07}%
A. Richard, J.-P. Comet, Necessary conditions for multistationarity in
discrete dynamical systems, {\emph{Discrete Appl. Math.}}, 155 (2007)
2403-2413.

\bibitem{R08}
A. Richard, An extension of a combinatorial fixed point theorem of
Shih and Dong, {\emph{Adv. Appl. Math.}}, 41 (2008) 620-627.  

\bibitem{R09}%
A. Richard, Positive circuits and maximal number of fixed points in
discrete dynamical systems, {\emph{Discrete Appl. Math.}},
(2009) in press.

\bibitem{R86}%
F. Robert, Discrete iterations: a metric study, in: Series in
Computational Mathematics, Vol. 6, Springer-Verlag,
Berlin-Heidelber-New York, 1986.

\bibitem{R95}%
F. Robert, Les syst\`emes dynamiques discrets, in: Math\'ematiques et
Applications, Vol.~19, Springer-Verlag, Berlin-Heidelber-New York,
1995.

\bibitem{SD05}%
M.-H. Shih and J.-L. Dong, A combinatorial analogue of the Jacobian
problem in automata networks, {\emph{Adv. Appl. Math.}}, 34 (2005)
30-46.

\bibitem{S89}%
E. H. Snoussi, Qualitative dynamics of a piecewise-linear differential
equations : a discrete mapping approach,
{\emph{Dynam. Stabil. Syst.}}, 4 (1989) 189-207.

\bibitem{ST93}%
E.H. Snoussi, R. Thomas, Logical identification of all steady states :
the concept of feedback loop caracteristic states,
{\emph{Bull. Math. Biol.}}, 55 (1993) 973-991.

\bibitem{S98}%
E.H. Snoussi, Necessary conditions for multistationarity and stable
periodicity, {\emph{J. Biol. Syst.}}, 6 (1998) 3-9.

\bibitem{S03}%
C. Soul\'e, Graphical requirements for multistationarity,
\emph{ComPlexUs}, 1 (2003) 123-133.

\bibitem{S06}%
C. Soul\'e, Mathematical approaches to differentiation and gene
regulation, \emph{C.R. Paris Biologies}, 329 (2006) 13-20.

\bibitem{T73}%
R. Thomas, Boolean formalization of genetic control
circuits, {\emph{J. Theor. Biol.}}, 42 (1973) 563-585.

\bibitem{T81}%
R. Thomas, On the relation between the logical structure of systems
and their ability to generate multiple steady states and sustained
oscillations, in: \emph{Series in Synergetics}, volume 9, pages
180-193, Springer, 1981.

\bibitem{TA90}%
R. Thomas, R. d'Ari, \emph{Biological Feedback}, CRC Press, 1990.

\bibitem{T91}%
R. Thomas, Regulatory Networks Seen as Asynchronous Automata : A
logical Description, {\emph{J. Theor. Biol.}}, 153
(1991) 1-23.

\bibitem{TK01}%
R. Thomas, M. Kaufman, Multistationarity, the basis of cell
differentiation and memory. I. \& II., \emph{Chaos}, 11 (2001)
170-195.

\end{thebibliography}
\end{document}